\documentclass[]{mn2e}
\usepackage{times}

\title[Physical conditions in the planetary nebula Abell 30]
{Physical conditions in the planetary nebula Abell 30}
\author[R. Wesson et al.]
{R. Wesson$^1$, X.-W. Liu$^1$, M. J. Barlow$^1$ \\
$^1$Department of Physics and Astronomy, University College London,
      Gower Street, London WC1E 6BT, UK\\}
\date{Received:}

\begin{document}
\maketitle

\begin{abstract}

We have analysed optical spectra of two of the hydrogen-deficient knots (J1 \& J3) in the born-again planetary nebula Abell~30, together with UV spectra of knots J3 \& J4.  We determine electron temperatures in the knots based on several diagnostics.  The [O~{\sc iii}] nebular-to-auroral transition ratio yields temperatures of the order of 17\,000~K.  The weak temperature dependence of the ratios of helium lines $\lambda$4471, $\lambda$5876 and $\lambda$6678 is used to derive a temperature of 8850~K for knot J3 and 4600~K for knot J1.  Ratios of O~{\sc ii} recombination lines, which directly measure the temperature in the coldest regions of the knots, are used to derive temperatures of 2500~K for knot J3, and just 500~K for knot J1.

We calculate abundances both from collisionally excited lines and from the well-observed recombination spectra of C, N, O and Ne ions.  The forbidden line abundances agree well with previous determinations, but the recombination line abundances are several hundred times higher.  These results confirm the scenario proposed by Harrington and Feibelman (1984) in which the knots contain a cold core highly enriched in heavy elements.  Forbidden lines are almost entirely emitted by the hot outer part of the knot, while recombination lines are emitted predominantly from the cold core.  The C/O ratios we derive imply that the knots are oxygen-rich, contrary to theoretical predictions for born-again nebulae (Iben et al 1983).

\end{abstract}

\begin{keywords}
ISM: abundances -- planetary nebulae: individual: Abell 30
\end{keywords}
%\nokeywords

\section{Introduction}

Abell~30 consists of a large ($\sim$ 120 arcsec across) spherical shell of low surface brightness, with several bright clumps of material concentrated within 10 arcsec of the central star.  The knots were first discovered by Jacoby (1979) and independently by Hazard et al (1980), and were found to be extremely hydrogen-deficient.  Knots J1 and J3 are collinear with the central star (Borkowski et al 1993), and according to Jacoby \& Chu (1989) are polar knots, while J2 and J4 are in the equatorial plane.  An evolutionary scenario to account for the hydrogen-deficient knots was proposed by Iben et al (1983), who suggested that in some cases, after the central star of a planetary nebula (CSPN) has become a white dwarf it might experience a final thermal pulse.  When this happens, most of the hydrogen left in the star is incorporated into the helium-burning shell and burned.

Two long-standing problems in the study of planetary nebulae are a) the discrepancy between ionic abundances measured from optical recombination lines (ORLs) and those measured from collisionally excited lines (CELs); and b) the discrepancy between temperatures measured from the hydrogen Balmer jump and those measured from [O~{\sc iii}] forbidden lines.  PN ORL abundances are found to be higher than CEL abundances by factors ranging from near unity to over 20 (see for example Liu et al 2000, 2001), while Balmer jump temperatures are invariably lower than [O~{\sc iii}] temperatures.  There is strong evidence that these two phenomena are related (Liu 2001, 2002).  In the most extreme case, Hf 2-2, ORL abundances are a factor of 80 higher than CEL abundances while T$_{BJ}$=T$_{[OIII]}$/10 (Liu 2002).  Several explanations have been offered to explain these discrepancies, including temperature fluctuations (Peimbert 1967), density fluctuations (Viegas \& Clegg 1994), abundance inhomogeneities (Torres-Peimbert, Peimbert \& Pe\~na 1990) and the presence of shock-waves (Peimbert, Sarmiento \& Fierro 1991).  In a recent study of the planetary nebula NGC\,6153, for which ORL abundances are a factor of 10 higher than CEL abundances, Liu et al (2000) suggested that the discrepancy might be caused by the presence of metal-rich knots within the nebula.  Empirical two-component models of NGC\,6153 were reasonably successful in reproducing the observed ORL and CEL line intensities, and also the observed Balmer jump and [O~{\sc iii}] forbidden line temperatures.  The posited knots in NGC\,6153 would be unseen due to the much higher surface brightness of the main nebula, but could be of similar origin to those in Abell~30.

Jacoby \& Ford (1983) determined ionic abundances within knots J3 and J4 of Abell~30, including a recombination line abundance for C$^{2+}$/H$^{+}$ from C~{\sc ii} $\lambda$4267 relative to H$\beta$.  The carbon abundance thus derived was almost half that of hydrogen, far higher than abundances of other heavy elements derived from CELs.  They considered this abundance questionable: Barker (1982) had earlier proposed that the $\lambda$4267 line in many nebulae might be enhanced by charge transfer, dielectronic recombination, line blending or resonance fluorescence.  Jacoby \& Ford also noted that such a huge carbon abundance would make it hard to maintain the high temperatures implied by standard diagnostics.  Later analysis of ultraviolet spectra of Abell~30 by Harrington \& Feibelman (1984) found that UV carbon CELs were very strong, implying that the carbon abundance is indeed high.  Their results suggested the presence of extreme temperature and abundance inhomogeneities within the knots of Abell~30.  From the carbon emission line ratios C~{\sc ii} $\lambda$4267/C~{\sc iii}] $\lambda$1908 and C~{\sc iii} $\lambda$4650/C~{\sc iv} $\lambda$1549, they found evidence for a much lower temperature in the knot than given by the standard [O~{\sc iii}] line ratios, and suggested that the very high abundances of CNO coolants within the knots had given rise to a very cool ($\sim$1000~K) but still ionised core.  Then, the ORL emission would come predominantly from the cool core, while the CEL emission would come from the much hotter outer regions of the knot.

Guerrero \& Manchado (1996) determined CEL abundances within knots J1-4, and found significantly higher helium abundances in the polar knots J1 and J3 than in the equatorial knots J2 and J4.  They established the rates of conversion of hydrogen into helium, and found that between 75 and 95 per cent of the original hydrogen has been burned into helium.  Oxygen and nitrogen abundances relative to the initial hydrogen abundance were found to be an order of magnitude lower than typical values in PNe.

In this work we present an analysis of long slit optical spectra and HST FOS ultraviolet spectra of knots J1 \& J3 of Abell~30.  We determine the temperature in the core of the knots directly, from the ratio of oxygen recombination lines, and confirm the existence of cold cores.  We determine abundances using both CELs and ORLs for carbon, nitrogen, oxygen and neon and show that the cold cores contain very high abundances of heavy elements.  The presence of cold CNO-rich cores in the knots of Abell~30 lends support to the scenario proposed by Liu et al (2000, 2001) and Liu (2002) to explain the abundance and temperature discrepancies in planetary nebulae.

\section{Spectroscopic observations}

\subsection{Optical spectra}

Abell~30 was observed using the double-armed ISIS spectrograph mounted on the 4.2m WHT at the Observatorio del Roque de los Muchachos, on La Palma, Spain, on the night of 20th February 2000.  The spectrograph slit was aligned such that the central star and the polar knots J1 and J3 were all observed.  The slit width used was 0.82 arcsec, which should have been sufficient to catch all the flux from the knots, which are approximately 0.3 arcsec across (Borkowski et al 1993).  Spectra covering wavelengths from 3400 to 5200 {\AA} and 5500 to 7000 {\AA} were taken.  Data were reduced using standard procedures in the {\sc midas}\footnote{{\sc midas} is developed and distributed by the European Southern Observatory.} package {\sc long92}.  They were bias subtracted, flat fielded, wavelength calibrated using a Cu-Ne lamp for the red spectra and a Cu-Ar lamp for the blue spectra.  The observations were flux calibrated by comparison with observations of the standard star Feige 34.

Hydrogen emission from the faint surrounding nebula is clearly visible in our long-slit spectra.  When measuring hydrogen emission from the knots, the contribution from the outer nebula was subtracted using a spectrum extracted from a region outside the knots covering the same number of rows on the CCD chip as the knots spectra.  The measured H$\beta$ flux from the knots was 1.24$\times$10$^{-16}$ erg/cm$^{2}$/s for J1 and 1.49$\times$10$^{-16}$ erg/cm$^{2}$/s for J3.  The contribution from the background for other lines is negligible.

Lines in the spectra of knots J1 \& J3 were identified and measured by fitting gaussian profiles.  A list of all the lines observed is given in Table~\ref{A30linelist}.  Line fluxes are conventionally normalised to H$\beta$=100 in nebular studies, but in this case as H$\beta$ is so weak, to avoid large numbers line fluxes are normalised such that I(H$\beta$)=1.

\setcounter{table}{1}
\begin{table*}
%\begin{minipage}{120mm}
\centering
\caption{Observed line fluxes}
\label{A30linelist}
\begin{tabular}{ccccccccccccc}

\hline

\multicolumn{3}{c}{Knot J1}&\multicolumn{3}{c}{Knot J3}& & & & & & & \\
$\lambda_{\rm obs}$&$F(\lambda)$&$I(\lambda)$&$\lambda_{\rm obs}$&$F(\lambda)$&$I(\lambda)$&Ion&$\lambda_{\rm 0}$&Mult&Lower Term&Upper Term&$g_1$&$g_2$\\

\hline

3426.37 & 21.14  & 33.49  & 3426.45 & 9.442  & 14.96  & [Ne V]  & 3425.86&F1     &  2p2 3P  &  2p2 1D & 5& 5\\
3444.32 & 3.395  & 5.335  & 3444.19 & 5.638  & 8.858  & O III   & 3444.07&V15    &  3p  3P  &  3d  3P*& 5& 5\\
3479.44 & 1.527  & 2.369  & *       & *      & *      & He I    & 3478.97&V43    &  2p  3P* & 15d  3D & 9&15\\
3635.09 & 0.636  & 0.934  & *       & *      & *      & He I    & 3634.25&V28    &  2p  3P* &  8d  3D & 9&15\\
3665.15 & 0.459  & 0.665  & *       & *      & *      & Ne II   & 3664.07&V1     &  3s  4P  &  3p  4P*& 6& 4\\
3694.99 & 0.502  & 0.718  & 3694.40 & 0.972  & 1.390  & Ne II   & 3694.21&V1     &  3s  4P  &  3p  4P*& 6& 6\\
*       & *      & *      & 3705.17 & 0.663  & 0.944  & He I    & 3705.02&V25    &  2p  3P* &  7d  3D & 9&15\\
3710.43 & 0.233  & 0.332  & *       & *      & *      & Ne II   & 3709.62&V1     &  3s  4P  &  3p  4P*& 4& 2\\
3713.79 & 0.790  & 1.121  & 3713.12 & 1.205  & 1.709  & Ne II   & 3713.08&V5     &  3s  2P  &  3p  2D*& 4& 6\\
3715.53 & 0.386  & 0.547  & *       & *      & *      & O III   & 3715.08&V14    &  3p  3P  &  3d  3D*& 5& 7\\
3726.77 & 9.235  & 13.03  & 3726.19 & 9.767  & 13.78  & [O II]  & 3726.03&F1     &  2p3 4S* &  2p3 2D*& 4& 4\\
3729.51 & 5.687  & 7.923  & 3728.92 & 5.735  & 7.990  & [O II]  & 3728.82&F1     &  2p3 4S* &  2p3 2D*& 4& 6\\
3750.48 & 0.578  & 0.807  & 3749.77 & 0.437  & 0.609  & O II    & 3749.48&V3     &  3s  4P  &  3p  4S*& 6& 4\\
3755.55 & 1.196  & 1.666  & 3754.91 & 0.989  & 1.378  & O III   & 3754.70&V2     &  3s  3P* &  3p  3D & 3& 5\\
*       & *      & *      & 3757.19 & 0.252  & 0.351  & O III   & 3757.24&V2     &  3s  3P* &  3p  3D & 1& 3\\
3760.65 & 1.694  & 2.354  & 3760.06 & 1.305  & 1.814  & O III   & 3759.87&V2     &  3s  3P* &  3p  3D & 5& 7\\
*       & *      & *      & 3777.41 & 0.382  & 0.526  & Ne II   & 3777.14&V1     &  3s  4P  &  3p  4P*& 2& 4\\
*       & *      & *      & 3819.89 & 0.802  & 1.086  & He I    & 3819.62&V22    &  2p  3P* &  6d  3D & 9&15\\
3834.75 & 0.584  & 0.785  & *       & *      & *      & He II   & 3834.89&4.18   &  4f+ 2F* & 18g+ 2G &32& *\\
3869.50 & 29.50  & 39.00  & 3868.92 & 26.83  & 35.46  & [Ne III]& 3868.75&F1     &  2p4 3P  &  2p4 1D & 5& 5\\
3889.42 & 5.123  & 6.707  & 3888.80 & 5.493  & 7.191  & He I    & 3888.65&He I   &  2s  3S &   2p  3P*& 9& 9\\
3924.16 & 0.552  & 0.713  & *       & *      & *      & He II   & 3923.48&4.15   &  4f+ 2F* & 15g+ 2G &32& *\\
3965.50 & 0.509  & 0.647  & 3964.85 & 0.493  & 0.627  & He I    & 3964.73&V5     &  2s  1S  &  4p  1P*& 1& 3\\
3968.25 & 9.027  & 11.47  & 3967.63 & 8.256  & 10.49  & [Ne III]& 3967.46&F1     &  2p4 3P  &  2p4 1D & 3& 5\\
4026.89 & 2.092  & 2.606  & 4026.31 & 1.902  & 2.370  & He I    & 4026.21&V18    &  2p  3P* &  5d  3D & 9&15\\
4042.17 & 0.624  & 0.775  & 4041.47 & 0.272  & 0.337  & N II    & 4041.31&V39b   &  3d  3F* & 4f 2[5] & 9&11\\
4060.14 & 0.600  & 0.744  & 4060.59 & 0.389  & 0.482  & O II    & 4060.60&V97    &  3d  2F  & 4f 2[4]*& 8& *\\
4068.68 & 0.411  & 0.508  & 4068.10 & 0.213  & 0.264  & C III   & 4067.94&V16    &  4f  3F* &  5g  3G & 5& 7\\
4069.66 & 0.538  & 0.665  & 4069.08 & 0.279  & 0.345  & C III   & 4068.92&V16    &  4f  3F* &  5g  3G & 7& 7\\
4070.37 & 0.737  & 0.910  & 4069.79 & 0.705  & 0.870  & O II    & 4069.62&V10    &  3p  4D* &  3d  4F & 2& 4\\
4070.63 & 1.179  & 1.457  & 4070.05 & 1.127  & 1.393  & O II    & 4069.89&V10    &  3p  4D* &  3d  4F & 4& 6\\
4071.01 & 0.702  & 0.868  & 4070.42 & 0.365  & 0.451  & C III   & 4070.26&V16    &  4f  3F* &  5g  3G & 9&11\\
4072.91 & 1.982  & 2.448  & 4072.32 & 1.515  & 1.871  & O II    & 4072.16&V10    &  3p  4D* &  3d  4F & 6& 8\\
4076.61 & 2.078  & 2.564  & 4076.03 & 1.511  & 1.864  & O II    & 4075.86&V10    &  3p  4D* &  3d  4F & 8&10\\
4079.59 & 0.310  & 0.382  & 4079.01 & 0.133  & 0.164  & O II    & 4078.84&V10    &  3p  4D* &  3d  4F & 4& 4\\
4084.65 & 0.340  & 0.418  & 4084.06 & 0.195  & 0.240  & O II    & 4083.90&V48b   &  3d  4F  &  4f  G4*& 6& 8\\
4085.86 & 0.306  & 0.378  & 4085.27 & 0.197  & 0.243  & O II    & 4085.11&V10    &  3p  4D* &  3d  4F & 6& 6\\
4087.90 & 0.328  & 0.404  & 4087.31 & 0.150  & 0.185  & O II    & 4087.15&V48c   &  3d  4F  &  4f  G3*& 4& 6\\
4090.04 & 1.042  & 1.283  & 4089.45 & 0.613  & 0.755  & O II    & 4089.29&V48a   &  3d  4F  &  4f  G5*&10&12\\
4098.15 & 2.355  & 2.893  & 4097.50 & 1.777  & 2.183  & O II    & 4097.26&V48b   &  3d  4F  &  4f  G4*& 8&10\\
*       & *      & *      & *       & *      & *      & O II    & 4097.25&V20    &  3p  4P* &  3d  4D & 2& 4\\
*       & *      & *      & *       & *      & *      & N III   & 4097.33&V1     &  3s  2S  &  3p  2P*& 2& 4\\
4103.98 & 1.002  & 1.231  & *       & *      & *      & He II   & 4100.04&4.12   &  4f+ 2F* & 12g+ 2G &32& *\\
*       & *      & *      & *       & *      & *      & H$\delta$     & 4101.73&H$\delta$ &  2p+ 2P* &  6d+ 2D & 8&72\\
4110.58 & 0.409  & 0.502  & *       & *      & *      & O II    & 4110.78&V20    &  3p  4P* &  3d  4D & 4& 2\\
4120.10 & 0.642  & 0.787  & 4119.42 & 0.388  & 0.475  & O II    & 4119.22&V20    &  3p  4P* &  3d  4D & 6& 8\\
4144.86 & 0.276  & 0.337  & 4144.39 & 0.500  & 0.611  & He I    & 4143.76&V53    &  2p  1P* &  6d  1D & 3& 5\\
4154.06 & 0.706  & 0.861  & *       & *      & *      & O II    & 4153.30&V19    &  3p  4P* &  3d  4P & 4& 6\\
*       & *      & *      & 4169.28 & 0.293  & 0.356  & O II    & 4169.22&V19    &  3p  4P* &  3d  4P & 6& 6\\
4187.85 & 0.273  & 0.332  & 4186.72 & 0.761  & 0.925  & C III   & 4186.90&V18    &  4f  1F* &  5g  1G & 7& 9\\
4200.51 & 0.435  & 0.528  & 4200.14 & 0.489  & 0.594  & He II   & 4199.83&4.11   &  4f+ 2F* & 11g+ 2G &32& *\\
*       & *      & *      & 4219.84 & 0.547  & 0.662  & Ne II   & 4219.37&V52a   &  3d  4D  & 4f 2[4]*& 8& 8\\
4237.72 & 0.380  & 0.459  & 4237.07 & 0.301  & 0.364  & N II    & 4236.91&V48a   &  3d  3D* & 4f 1[3] & 3& 5\\
4242.11 & 0.435  & 0.525  & *       & *      & *      & N II    & 4241.78&V48b   &  3d  3D* & 4f 1[4] & 7& 9\\
4267.95 & 5.693  & 6.851  & 4267.36 & 4.699  & 5.655  & C II    & 4267.15&V6     &  3d  2D  &  4f  2F*&10&14\\
4276.37 & 0.642  & 0.772  & 4275.92 & 1.056  & 1.270  & O II    & 4276.28&V67b   &  3d  4D  &  4f  F3*& 6& 6\\
*       & *      & *      & *       & *      & *      & O II    & 4276.62&V53c   &  3d  4P  &  4f  D1*& 4& 4\\
*       & *      & *      & *       & *      & *      & O II    & 4276.71&V53c   &  3d  4P  &  4f  D1*& 4& 2\\
*       & *      & *      & *       & *      & *      & O II    & 4276.75&V67b   &  3d  4D  &  4f  F3*& 6& 8\\
4277.92 & 0.846  & 1.017  & 4277.75 & 0.528  & 0.635  & O II    & 4277.43&V67c   &  3d  4D  &  4f  F2*& 2& 4\\
*       & *      & *      & *       & *      & *      & O II    & 4277.89&V67b   &  3d  4D  &  4f  F3*& 8& 8\\
4282.10 & 0.043  & 0.052  & 4281.60 & 0.103  & 0.125  & O II    & 4281.32&V53b   &  3d  4P  &  4f  D2*& 6& 6\\
4283.76 & 0.472  & 0.567  & 4283.24 & 0.329  & 0.399  & O II    & 4282.96&V67c   &  3d  4D  &  4f  F2*& 4& 6\\
4284.53 & 0.028  & 0.034  & 4284.01 & 0.253  & 0.307  & O II    & 4283.73&V67c   &  3d  4D  &  4f  F2*& 4& 4\\

\end{tabular}
%\end{minipage}
\end{table*}

\setcounter{table}{1}
\begin{table*}
%\begin{minipage}{120mm}
\centering
\caption{{\it --continued}}
\begin{tabular}{ccccccccccccc}

\hline

\multicolumn{3}{c}{Knot J1}&\multicolumn{3}{c}{Knot J3}& & & & & & & \\
$\lambda_{\rm obs}$&$F(\lambda)$&$I(\lambda)$&$\lambda_{\rm obs}$&$F(\lambda)$&$I(\lambda)$&Ion&$\lambda_{\rm 0}$&Mult&Lower Term&Upper Term&$g_1$&$g_2$\\

\hline

4286.49 & 0.352  & 0.423  & 4285.97 & 0.232  & 0.281  & O II    & 4285.69&V78b   &  3d  2F  &  4f  F3*& 6& 8\\
4292.06 & 0.366  & 0.438  & 4291.52 & 0.239  & 0.268  & O II    & 4291.25&V55    &  3d  4P  &  4f  G3*& 6& 8\\
4293.01 & 0.071  & 0.085  & 4292.49 & 0.420  & 0.508  & O II    & 4292.21&V78c   &  3d  2F  &  4f  F2*& 6& 6\\
4295.61 & 0.456  & 0.547  & 4295.09 & 0.551  & 0.666  & O II    & 4294.78&V53b   &  3d  4P  &  4f  D2*& 4& 6\\
4304.53 & 0.940  & 1.127  & 4303.90 & 0.933  & 1.118  & O II    & 4303.82&V53a   &  3d  4P  &  4f  D3*& 6& 8\\
4318.02 & 0.461  & 0.551  & 4317.59 & 0.366  & 0.438  & O II    & 4317.14&V2     &  3s  4P  &  3p  4P*& 2& 4\\
4320.31 & 0.345  & 0.412  & 4319.86 & 0.179  & 0.213  & O II    & 4319.63&V2     &  3s  4P  &  3p  4P*& 4& 6\\
*       & *      & *      & 4325.91 & 0.147  & 0.175  & O II    & 4325.76&V2     &  3s  4P  &  3p  4P*& 2& 2\\
4339.50 & 0.824  & 0.982  & 4338.87 & 0.573  & 0.682  & He II   & 4338.67&4.10   &  4f+ 2F* & 10g+ 2G &32& *\\
4341.28 & 1.469  & 1.751  & 4340.55 & 0.547  & 0.652  & H$\gamma$& 4340.47&H$\gamma$&  2p+ 2P* &  5d+ 2D & 8&50\\
4343.01 & 0.492  & 0.586  & 4342.24 & 0.540  & 0.644  & O II    & 4342.00&V77    &  3d 2F   & 4f 2[5]*& 8&10\\
4346.08 & 0.778  & 0.927  & 4345.64 & 0.175  & 0.208  & O II    & 4345.56&V2     &  3s  4P  &  3p  4P*& 4& 2\\
*       & *      & *      & *       & *      & *      & O II    & 4345.55&V65c   &  3d  4D  &  4f  G3*& 8& 8\\
4350.31 & 0.685  & 0.815  & 4349.61 & 0.622  & 0.740  & O II    & 4349.43&V2     &  3s  4P  &  3p  4P*& 6& 6\\
4364.02 & 4.616  & 5.486  & 4363.38 & 3.668  & 4.360  & [O III] & 4363.21&F2     &  2p2 1D  &  2p2 1S & 5& 1\\
4367.90 & 1.299  & 1.544  & 4366.99 & 0.579  & 0.688  & O II    & 4366.89&V2     &  3s  4P  &  3p  4P*& 6& 4\\
4370.52 & 0.592  & 0.703  & *       & *      & *      & Ne II   & 4369.86&V56    &  3d  4F  & 4f 0[3]*& 4& 6\\
4372.70 & 0.315  & 0.373  & *       & *      & *      & O II    & 4371.62&V76b   &  3d  2F  &  4f  G4*& 8&10\\
4380.12 & 2.207  & 2.615  & 4379.48 & 1.652  & 1.958  & N III   & 4379.11&V18    &  4f  2F* &  5g  2G &14&18\\
4388.83 & 0.488  & 0.577  & 4387.90 & 0.478  & 0.555  & He I    & 4387.93&V51    &  2p  1P* &  5d  1D & 3& 5\\
4392.85 & 0.686  & 0.812  & 4391.99 & 0.311  & 0.369  & Ne II   & 4391.99&V55e   &  3d  4F  & 4f 2[5]*&10&12\\
*       & *      & *      & *       & *      & *      & Ne II   & 4392.00&V55e   &  3d  4F  & 4f 2[5]*&10&10\\
4398.84 & 0.371  & 0.439  & 4398.62 & 0.293  & 0.347  & Ne II   & 4397.99&V57b   &  3d  4F  & 4f 1[4]*& 6& 8\\
4409.91 & 0.759  & 0.896  & 4409.19 & 0.777  & 0.917  & Ne II   & 4409.30&V55e   &  3d  4F  & 4f 2[5]*& 8&10\\
4413.79 & 0.326  & 0.385  & 4414.85 & 0.533  & 0.629  & Ne II   & 4413.22&V65    &  3d  4P  & 4f 0[3]*& 6& 8\\
4415.83 & 0.561  & 0.662  & *       & *      & *      & O II    & 4414.90&V5     &  3s  2P  &  3p  2D*& 4& 6\\
4417.83 & 0.393  & 0.464  & 4417.17 & 0.361  & 0.426  & O II    & 4416.97&V5     &  3s  2P  &  3p  2D*& 2& 4\\
4429.46 & 0.437  & 0.515  & 4428.91 & 0.512  & 0.603  & Ne II   & 4428.64&V60c   &  3d  2F  & 4f 1[3]*& 6& 8\\
*       & *      & *      & *       & *      & *      & Ne II   & 4428.52&V61b   &  3d  2D  & 4f 2[3]*& 6& 8\\
4432.09 & 0.205  & 0.242  & 4431.46 & 0.390  & 0.459  & Ne II   & 4430.94&V61a   &  3d  2D  & 4f 2[4]*& 6& 8\\
*       & *      & *      & 4434.82 & 0.599  & 0.704  & N I ?   & 4435.11&       &  3p  2P* &  3d  2P & 2& 2\\
4442.55 & 0.118  & 0.139  & *       & *      & *      & N II    & 4442.02&V55a   &  3d  3P* & 4f 2[3] & 3& 5\\
4448.60 & 0.314  & 0.367  & 4448.26 & 0.355  & 0.416  & O II    & 4448.19&V35    &  3p' 2F* &  3d' 2F & 8& 8\\
4454.25 & 0.601  & 0.702  & *       & *      & *      &         &        &       &          &         &  &  \\
4459.62 & 0.398  & 0.464  & *       & *      & *      & N II    & 4459.94&V21    &  3p  3D  &  3d  3P*& 3& 1\\
4467.05 & 0.307  & 0.357  & 4466.60 & 0.247  & 0.287  & O II    & 4466.42&V86b   &  3d  2P  &  4f  D2*& 4& 6\\
4472.33 & 3.934  & 4.567  & 4471.67 & 3.902  & 4.530  & He I    & 4471.50&V14    &  2p  3P* &  4d  3D & 9&15\\
4476.39 & 0.375  & 0.434  & *       & *      & *      &         &        &       &          &         &  &  \\
*       & *      & *      & 4491.53 & 0.238  & 0.274  & O II    & 4491.23&V86a   &  3d  2P  &  4f  D3*& 4& 6\\
4531.98 & 0.233  & 0.265  & 4530.71 & 0.293  & 0.333  & N II    & 4530.41&V58b   &  3d  1F* & 4f 2[5] & 7& 9\\
4542.51 & 0.813  & 0.920  & 4541.82 & 0.841  & 0.952  & He II   & 4541.59&4.9    &  4f+ 2F* &  9g+ 2G &32& *\\
*       & *      & *      & 4602.60 & 0.623  & 0.690  & O II    & 4602.13&V92b   &  3d  2D  &  4f  F3*& 4& 6\\
4610.61 & 0.838  & 0.925  & 4609.70 & 0.379  & 0.419  & O II    & 4609.44&V92a   &  3d  2D  &  4f  F4*& 6& 8\\
*       & *      & *      & 4624.93 & 0.265  & 0.291  & [Ar V]  & 4625.53&       &  3p2 1D  &  3p2 1S & 5& 1\\
4635.02 & 0.363  & 0.397  & 4634.27 & 0.194  & 0.212  & N III   & 4634.14&V2     &  3p  2P* &  3d  2D & 2& 4\\
4640.24 & 1.055  & 1.153  & 4638.99 & 0.630  & 0.689  & O II    & 4638.86&V1     &  3s  4P  &  3p  4D*& 2& 4\\
*       & *      & *      & 4640.77 & 0.544  & 0.594  & N III   & 4640.64&V2     &  3p  2P* &  3d  2D & 4& 6\\
4642.48 & 2.150  & 2.347  & 4641.94 & 1.479  & 1.615  & O II    & 4641.81&V1     &  3s  4P  &  3p  4D*& 4& 6\\
*       & *      & *      & 4641.97 & 0.039  & 0.042  & N III   & 4641.84&V2     &  3p  2P* &  3d  2D & 4& 4\\
4649.00 & 0.474  & 0.517  & 4647.55 & 0.352  & 0.384  & C III   & 4647.42&V1     &  3s  3S  &  3p  3P*& 3& 5\\
4650.04 & 2.560  & 2.787  & 4649.26 & 2.306  & 2.510  & O II    & 4649.13&V1     &  3s  4P  &  3p  4D*& 6& 8\\
4651.59 & 0.761  & 0.829  & 4650.38 & 0.211  & 0.230  & C III   & 4650.25&V1     &  3s  3S  &  3p  3P*& 3& 3\\
4651.83 & 0.158  & 0.172  & 4650.97 & 0.677  & 0.737  & O II    & 4650.84&V1     &  3s  4P  &  3p  4D*& 2& 2\\
4653.06 & 0.095  & 0.103  & 4651.60 & 0.070  & 0.077  & C III   & 4651.47&V1     &  3s  3S  &  3p  3P*& 3& 1\\
4659.15 & 0.302  & 0.327  & 4658.61 & 0.136  & 0.147  & C IV    & 4658.64&V8     &  5f  2F* &  6g  2G &14&18\\
4662.39 & 0.932  & 1.010  & 4661.75 & 0.778  & 0.843  & O II    & 4661.63&V1     &  3s  4P  &  3p  4D*& 4& 4\\
*       & *      & *      & 4669.40 & 0.178  & 0.193  & O II    & 4669.27&V89b   &  3d  2D  &  4f  D2*& 4& 6\\
*       & *      & *      & 4673.87 & 0.116  & 0.125  & O II    & 4673.73&V1     &  3s  4P  &  3p  4D*& 4& 2\\
4677.04 & 0.661  & 0.713  & 4676.38 & 0.594  & 0.641  & O II    & 4676.24&V1     &  3s  4P  &  3p  4D*& 6& 6\\
4686.59 & 34.92  & 37.63  & 4685.92 & 28.93  & 31.17  & He II   & 4685.68&3.4    &  3d+ 2D  &  4f+ 2F*&18&32\\
4713.27 & 2.017  & 2.165  & 4711.35 & 0.388  & 0.416  & He I    & 4713.17&V12    &  2p  3P* &  4s  3S & 9& 3\\
*       & *      & *      & *       & *      & *      & [Ar IV] & 4711.37&F1     &  3p3 4S* &  3p3 2D*& 4& 6\\
4715.08 & 1.409  & 1.498  & 4714.97 & 1.883  & 2.002  & [Ne IV] & 4714.17&F1     &  2p3 2D* &  2p3 2P*& 6& 4\\
*       & *      & *      & *       & *      & *      & [Ne IV] & 4715.66&F1     &  2p3 2D* &  2p3 2P*& 6& 2\\

\end{tabular}
%\end{minipage}
\end{table*}

\setcounter{table}{1}
\begin{table*}
%\begin{minipage}{120mm}
\centering
\caption{{\it --continued}}
\begin{tabular}{ccccccccccccc}

\hline

\multicolumn{3}{c}{Knot J1}&\multicolumn{3}{c}{Knot J3}& & & & & & & \\
$\lambda_{\rm obs}$&$F(\lambda)$&$I(\lambda)$&$\lambda_{\rm obs}$&$F(\lambda)$&$I(\lambda)$&Ion&$\lambda_{\rm 0}$&Mult&Lower Term&Upper Term&$g_1$&$g_2$\\

\hline

4725.48 & 4.111  & 4.363  & 4725.47 & 1.995  & 2.118  & [Ne IV] & 4724.15&F1     &  2p3 2D* &  2p3 2P*& 4& 4\\
*       & *      & *      & *       & *      & *      & [Ne IV] & 4725.62&F1     &  2p3 2D* &  2p3 2P*& 4& 2\\
4741.10 & 0.634  & 0.670  & 4740.55 & 0.308  & 0.325  & [Ar IV] & 4740.17&F1     &  3p3 4S* &  3p3 2D*& 4& 4\\
4860.17 & 1.253  & 1.318  & 4859.49 & 1.243  & 1.307  & He II   & 4859.32&4.8    &  4f+ 2F* &  8g+ 2G &32& *\\
4862.35 & 1.000  & 1.000  & 4861.41 & 1.000  & 1.000  & H$\beta$& 4861.33&H$\beta$& 2p+ 2P* &  4d+ 2D & 8&32\\
4922.85 & 1.164  & 1.169  & 4922.08 & 1.217  & 1.222  & He I    & 4921.93&V48    &  2p  1P* &  4d  1D & 3& 5\\
4925.36 & 0.258  & 0.253  & 4925.05 & 0.332  & 0.326  & O II    & 4924.53&V28    &  3p  4S* &  3d  4P & 4& 6\\
4959.84 & 67.03  & 65.71  & 4959.07 & 61.59  & 60.38  & [O III] & 4958.91&F1     &  2p2 3P  &  2p2 1D & 3& 5\\
*       & *      & *      & 4995.23 & 0.295  & 0.285  & N II    & 4994.36&V24    &  3p  3S  &  3d  3P*& 3& 3\\
5007.76 & 203.2  & 193.1  & 5006.99 & 186.7  & 177.4  & [O III] & 5006.84&F1     &  2p2 3P  &  2p2 1D & 5& 5\\
5016.56 & 1.384  & 1.310  & 5015.83 & 1.277  & 1.209  & He I    & 5015.68&V4     &  2s  1S  &  3p  1P*& *& 1\\
*       & *      & *      & 5664.84 & 0.299  & 0.243  & N II    & 5666.63&V3     &  3s  3P* &  3p  3D & 3& 5\\
*       & 0.691  & 0.563  & 5677.83 & 0.509  & 0.415  & N II    & 5679.56&V3     &  3s  3P* &  3p  3D & 5& 7\\
*       & *      & *      & 5753.64 & 0.374  & 0.302  & [N II]  & 5754.60&F3     &  2p2 1D  &  2p2 1S & 5& 1\\
5874.75 & 17.62  & 13.97  & 5874.16 & 15.77  & 12.50  & He I    & 5875.66&V11    &  2p  3P* &  3d  3D & 9&15\\
*       & *      & *      & 6298.79 & 1.196  & 0.888  & [O I]   & 6300.34&F1     &  2p4 3P  &  2p4 1D & 5& 5\\
*       & *      & *      & 6362.26 & 0.368  & 0.270  & [O I]   & 6363.78&F1     &  2p4 3P  &  2p4 1D & 3& 5\\
*       & *      & *      & 6460.54 & 0.696  & 0.502  & C II    & 6461.95&       &  4f  2F* &  6g  2G &14&18\\
6547.38 & 5.911  & 4.202  & 6546.69 & 5.050  & 3.590  & [N II]  & 6548.10&F1     &  2p2 3P  &  2p2 1D & 3& 5\\
6560.47 & 9.984  & 7.078  & 6559.66 & 7.886  & 5.591  & He II   & 6560.10&4.6    &  4f+ 2F* &  6g+ 2G &32& *\\
*       & *      & *      & *       & *      & *      & H$\alpha$& 6562.77&H$\alpha$&  2p+ 2P* &  3d+ 2D & 8&18\\
6582.59 & 17.70  & 12.49  & 6581.94 & 14.68  & 10.36  & [N II]  & 6583.50&F1     &  2p2 3P  &  2p2 1D & 5& 5\\
6677.32 & 5.450  & 3.785  & 6676.67 & 5.068  & 3.519  & He I    & 6678.16&V46    &  2p  1P* &  3d  1D & 3& 5\\
*       & *      & *      & 6683.17 & 0.218  & 0.151  & He II   & 6683.20&5.13   &  5g+ 2G  & 13h+ 2H*&50& *\\

\hline

\end{tabular}
%\end{minipage}
\end{table*}

\subsection{Ultraviolet spectra}

Knots J3 \& J4 of Abell~30 were observed by the Faint Object Spectrograph aboard the Hubble Space Telescope on December 8th 1994, as part of GO program 5690, and these observations were obtained by us from the HST archive.  Observations were taken using gratings G130H, G190H and G270H, which together cover wavelengths from 1100 to 3300 {\AA}.  The spectral resolutions are 1 {\AA}, 1.47 {\AA} and 2.09 {\AA} respectively.  All observations were taken using a round aperture with diameter 0.86 arcsec, which should have captured all the flux from the knots, based on their dimensions given by Borkowski et al (1993).  Standard pipeline reduction yields FITS files containing a wavelength array and a flux array, and these were combined in {\sc midas}.

Two exposures were taken with each grating, and these were combined, except for G190H on J3, for which there is only one exposure.  The observations taken using grating G130H were rebinned by a factor of five to improve the signal to noise ratio, while those using grating G190H were rebinned by a factor of two.  Observations using grating G270H were not rebinned.  Lines were identified and measured using {\sc midas} by fitting gaussian profiles.  A list of all lines is given in Table~\ref{A30FOSlinelist}.  The observed fluxes are given in columns 2 (J3) and 5 (J4), while column 3 gives the fluxes from knot J3 dereddened and normalised to H$\beta$=1, from the optical spectra.

A strong, broad feature at $\sim$2140 {\AA} which appears in the UV spectra of both knots is probably due to sky emission (Lyons et al 1993).  The [O~{\sc ii}] line at $\lambda$2470 is also affected by sky emission, being much stronger than would be expected given the O$^{+}$/H$^{+}$ abundance implied by the $\lambda\lambda$3726,3729 optical lines.

\setcounter{table}{1}
\begin{table*}
%\begin{minipage}{120mm}
\centering
\caption{HST FOS line fluxes}
\label{A30FOSlinelist}
\begin{tabular}{ccccccccc}

\hline

\multicolumn{4}{c}{Knot J3}&\multicolumn{3}{c}{Knot J4}& & \\
$\lambda_{\rm obs}$&$F(\lambda)$&$I(\lambda)$&$\lambda_{\rm obs}$&$F(\lambda)$&Ion&$\lambda_{\rm 0}$\\
                   & (10$^{-16}$erg\,cm$^{-2}$\,s$^{-1}$) &($I({\rm H}\beta) = 1$) & & (10$^{-16}$erg\,cm$^{-2}$\,s$^{-1}$) & & \\
\hline

%1401.71 & 26   & 37     & 1400.89 & 42  & O IV]  & 1401.16 \\              
%1406.14 & 24   & 34     & 1407.65 & 27  & O IV]  & 1404.81 \\              Sky emission?  Harrington (1996) doesn't mention them
1483.60 & 24   & 32     & *       &  *  & N IV]  & 1483.32 / 1486.50 \\              
1548.57 & 44   & 57     & 1548.47 & 77  & C IV   & 1548.20 \\              
1551.62 & 45   & 58     & 1552.71 & 49  & C IV   & 1550.78 \\              
1640.00 & 170  & 210    & 1639.10 & 38  & He II  & 1640.42 \\              
 * &       *   & *      & 1772.23 & 4.3 & ?      &      ?  \\              
1905.95 & 24   & 37     & 1905.44 & 21  & C III] & 1906.68 \\              
1908.28 & 23   & 35     & 1907.64 & 10  & C III] & 1908.73 \\              
2297.86 & 5.7  & 12     & *       & *   & C III  & 2297.58 \\              
2423.42 & 84   & 190    & 2422.74 & 70  & [Ne IV]& 2422.51 / 2425.15 \\    
2470.85 & 6.2  & 14     & *       & *   & [O II] & 2471.05 / 2471.12 \\    
2511.74 & 4.3  & 9.5    & *       & *   & He II  & 2511.96 \\              
2733.31 & 6.0  & 12     & *       & *   & He II  & 2734.11 \\              
2836.54 & 4.8  & 8.7    & *       & *   & O III  & 2836.31 \\              
3048.26 & 3.8  & 5.6    & *       & *   & O III  & 3047.10 \\              
3133.89 & 15   & 20     & 3132.52 & 1.6 & O III  & 3132.79 \\              
3188.75 & 1.6  & 2      & *       & *   & He I   & 3187.74 \\              
3203.75 & 9.1  & 11     & 3202.57 & 1.9 & He II  & 3203.10 \\              

\hline

\end{tabular}
%\end{minipage}
\end{table*}

\section{Nebular Analysis}

\subsection{Extinction}

Greenstein (1980) first noted the unusual extinction towards Abell~30.  Instead of the usual 2200 {\AA} feature in the extinction curve, increased extinction is seen at 2600 {\AA}, and UV extinction is much lower than implied by a standard galactic curve.  Greenstein determined an extinction curve by dereddening spectra of the central star to match a black body of 200\,000~K.  Later, Jeffery (1995) found a similar curve by dereddening to match a model atmosphere with a temperature of 114\,000~K.  Harrington (1996) followed an approach which avoids the uncertainties of model comparisons, by instead comparing the spectrum of the central star of Abell~30 with a dereddened spectrum of the central star of Abell~78, which has very similar properties to those of A30, but has normal interstellar extinction.  Over their common spectral range, Harrington's curve is very similar to that of Greenstein.  Greenstein's curve covers visual wavelengths as well as ultraviolet, therefore we use it to deredden both our optical and UV line fluxes.

Previous studies have suggested that while the spectrum of the central star is reddened, the knots are not (Guerrero \& Manchado 1996).  We have measured the extinction in the knots using hydrogen and helium line ratios, and we find that extinction in the knots is comparable to that of the central star.  The use of appropriate directly measured temperatures when calculating theoretical line ratios results in consistent determination of c(H$\beta$) from all the diagnostics used.

The extinction to the central star was determined by dereddening its optical spectrum with the Greenstein extinction curve to match the continuum emission of a black body with a temperature of $10^{5}$K.  $c(H\beta)$ was found from this method to be 0.60, which agrees fairly well with previous determinations.  Cohen et al (1977) derived c(H$\beta$)=0.44, while Jeffery (1995) found A$_{\rm v}$=1.18 for the central star, which is equivalent to c(H$\beta$)=0.61.

The extinction to the knots was estimated from I(H$\alpha$)/I(H$\beta$).  In our spectra, the He~{\sc ii} Pickering series line at $\lambda$4860 is resolved from H$\beta$, but in the lower resolution red spectra, H$\alpha$ is blended with He~{\sc ii} $\lambda$6560 emission.  This is corrected for using the observed flux of the He~{\sc ii} 4686 line -- at the temperatures derived from helium lines in the knots (see section~\ref{tempsanddensities}), I($\lambda$6560)=0.13I($\lambda$4686) (Storey and Hummer 1995).  After subtraction of the helium contribution, H$\alpha$/H$\beta$ is found to be 5.44 (J1) and 4.13 (J3).  The intrinsic H$\alpha$/H$\beta$ ratio depends on the temperature, and adopting the electron temperature determined from helium line ratios in the following section, i.e. 4600~K in J1 and 8850~K in J3, gives $c(H\beta)$=1.02 to J1 and 0.64 to J3.  Using the lower temperatures given by O~{\sc ii} recombination lines, $c(H\beta)$=0.60 in J1 and 0.45 in J3.

For J3, the UV observations allow a determination of $c(H\beta)$ from the He~{\sc ii} $\lambda$1640/$\lambda$4686 ratio.  The observed value is 3.94, while the predicted value at 8850~K (the temperature derived below from He~{\sc i} lines) is 6.40, giving $c(H\beta)$=0.55, which agrees well with the value derived from the other diagnostics.

We adopt c(H$\beta$)=0.60 to deredden the spectra of both knots.

\subsection{Temperatures and densities}
\label{tempsanddensities}

The temperature structure of the knots in A30 has been suggested to be complex.  To explain the observed ratios of carbon ORLs to CELs, Harrington and Feibelman (1984) proposed cool, carbon-rich cores in the knots.  The large abundances of CNO coolants relative to He would result in material which is very cool ($\sim$1000~K) but still highly ionised, as the heat input by photoionisation would be balanced by radiation in the infrared fine-structure lines of the heavy elements.  We have used several temperature diagnostics to study the temperature structure of knots J1 \& J3.

The electron densities of the knots were measured from the [O~{\sc ii}] lines at $\lambda\lambda$3726,3729, and found to be 2800 cm$^{-3}$ and 3200 cm$^{-3}$ for knots J1 and J3 respectively.  Adopting this density, the [O~{\sc iii}] ($\lambda$4959 + $\lambda$5007)/$\lambda$4363 ratio gives temperatures of 17\,960~K and 16\,680~K respectively.

Lines of He~{\sc i} are well observed in our long slit spectra.  The ratios of the strong lines at $\lambda$4471, $\lambda$5876 and $\lambda$6678 are moderately dependent on temperature: $\lambda$5876/$\lambda$4471 varies from $\sim$2.5 at 20\,000~K to $\sim$3.5 at 5000~K, while $\lambda$6678/$\lambda$4471 varies from $\sim$0.6 to $\sim$0.9 over the same temperature range (Smits 1996).  Collisional excitation from the metastable 2s\,$^3$S level can be important in enhancing the intensities of these lines, and the predicted ratios given in Smits (1996) were corrected for this effect using formulae from Kingdon \& Ferland (1995), which are derived using a 29 state quantal calculation of He~{\sc i} extending to $n$=5.  We use the average of the temperatures implied by these two line ratios to derive temperatures of 4600~K for knot J1 and 8850~K for knot J3.  Although the lines are well detected, with fluxes accurate to within 5 per cent, the shallow slope of the temperature dependence of the line ratios means the error on these temperature measurements is of the order of $\pm$2000~K.

Some O~{\sc ii} recombination line ratios are temperature-sensitive.  For example, the ratio of the lines at $\lambda$4649 and $\lambda$4089 varies from approximately 2 to 6 between 1000 and 20\,000~K (Storey 1994, Liu et al 1995).  These lines are weak but fairly well detected in our spectra, with errors on the flux of the order of 5 per cent for $\lambda$4649 and 20 per cent for $\lambda$4089.  We use their ratio to derive temperatures of 500~K for J1 and 2100~K for J3.  The ratio of the lines at $\lambda$4075 and $\lambda$4089 varies by a factor of two between 500~K and 20\,000~K, and the observed values of this ratio (given in Table~\ref{temperatures}) also imply very low temperatures, of 400~K for J1 and 2800~K for J3.  The errors on these values are of the order of $\pm$ 2000~K.

The three O~{\sc ii} lines at $\lambda$4089, $\lambda$4075 and $\lambda$4649 are the highest J-value quartet transitions from the 3d-4f, 3p-3d and 3s-3p configurations respectively.  Their emission can only be produced by recombination from the 2p$^{2}$\,$^{3}$P$_{2}$ level of O$^{2+}$.  Other O~{\sc ii} line ratios which are temperature-sensitive include $\lambda$4072/$\lambda$4089 and $\lambda$4414/$\lambda$4089.  However, the observed values of these ratios imply much higher temperatures, of the order of 5-20~kK.  Liu (2002) observes a similar phenomenon for other nebulae, and suggests that one possibility is that the 2p$^{2}$\,$^{3}$P$_{2}$ level of the O$^{2+}$ ion is underpopulated relative to its statistical equilibrium value.  Temperatures measured from two lines which both originate via recombination from this level would not be affected, but temperatures measured from two lines which originate via recombinations from different levels would then be unreliable.

The different temperatures derived here from forbidden lines, He~{\sc i} recombination lines and O~{\sc ii} recombination lines are exactly what would be expected if the knots contain very cool, CNO-rich, ionised cores.  Essentially all CEL emission would come from the warmer outer layers, with most of the ORL emission coming from the cooler cores.  Helium emission would come from both components if helium is not enriched in the core.  Therefore, in the presence of a cold ionised core one would expect to find that T$_{{\rm [OIII]}} > $T$_{{\rm He I}} > $T$_{{\rm OII}}$ (Liu 2002).  The derived temperatures follow this relation.

\begin{table}
%\begin{minipage}{120mm}
\centering
\caption{Derived electron temperatures in Abell~30}
\label{temperatures}
\begin{tabular}{llcc}

\hline
      & Lines  & Ratio  & Temperature(K) \\
\hline
J1    & [O~{\sc iii}] ($\lambda\lambda$4959+5007)/$\lambda$4363 & 47.18 & 17960 \\
      & He~{\sc i} ($\lambda5876$/$\lambda$4471)               & 3.059 & 4900  \\
      & He~{\sc i} ($\lambda6678$/$\lambda$4471)               & 0.829 & 4300  \\
      & O~{\sc ii} ($\lambda4649$/$\lambda$4089)               & 2.172 & 500   \\
      & O~{\sc ii} ($\lambda4075$/$\lambda$4089)               & 1.998 & 400   \\
J3    & [O~{\sc iii}] ($\lambda\lambda$4959+5007)/$\lambda$4363 & 54.54 & 16680 \\
      & He~{\sc i} ($\lambda5876$/$\lambda$4471)               & 2.759 & 9240  \\
      & He~{\sc i} ($\lambda6678$/$\lambda$4471)               & 0.777 & 8450  \\
      & O~{\sc ii} ($\lambda4649$/$\lambda$4089)               & 3.325 & 2100  \\
      & O~{\sc ii} ($\lambda4075$/$\lambda$4089)               & 2.470 & 2800   \\
\hline

\end{tabular}
\end{table}

\subsection{Ionic abundances from CELs}

Abundances relative to hydrogen were derived from CELs for ionic species of C, N, O and Ne.  The temperature implied by the [O~{\sc iii}] nebular to auroral line ratio and the density implied by the [O~{\sc ii}] $\lambda$3726/$\lambda$3729 ratio were adopted for all species in both knots.  Transition probabilities and collision strengths were taken from the references listed in Table~\ref{CELrefs}.  The results are given in Table~\ref{CELionicabunds}.  For species in common, the results agree fairly well with the abundances previously derived for J1 and J3 by Guerrero \& Manchado (1996).

%The abundance of O~{\sc iv} derived for knot J3 from the FOS spectra is extremely high, at 1.75$\times$10$^{2}$.  Not much oxygen would be expected in the form of O$^{3+}$ and so the strength of these lines may imply the presence of shock waves in the knot.  Evidence for shocks in the hydrogen-deficient knot of Abell~58 was found by Guerrero \& Manchado (1996).

\begin{table}
%\begin{minipage}{120mm}
\centering
\caption{Ionic abundances from CELs}
\label{CELionicabunds}
\begin{tabular}{ccccc}

\hline
X$^{i+}$/H$^{+}$ & Lines & J1 & J3 \\
\hline

C$^{2+}$/H$^{+}$  & $\lambda\lambda$1906,1908 &          & 7.01(-4)\\
C$^{3+}$/H$^{+}$  & $\lambda\lambda$1548,1551 &          & 7.66(-4)\\
N$^{+}$/H$^{+}$   & $\lambda\lambda$6548,6584 & 6.52(-5) & 6.24(-5)\\
N$^{3+}$/H$^{+}$  & $\lambda\lambda$1483,1486 &          & 4.66(-4)\\
O$^{+}$/H$^{+}$   & $\lambda\lambda$3726,3729 & 1.57(-4) & 2.05(-4)\\
O$^{2+}$/H$^{+}$  & $\lambda\lambda$4959,5007 & 1.41(-3) & 1.52(-3)\\
O$^{3+}$/H$^{+}$  & $\lambda\lambda$1401,1404 &          & 1.75(-2)\\
Ne$^{2+}$/H$^{+}$ & $\lambda\lambda$3868,3967 & 5.95(-4) & 6.55(-4)\\
Ne$^{3+}$/H$^{+}$ & $\lambda\lambda$4715,4726 & 3.97(-3) & 5.12(-3)\\
Ne$^{3+}$/H$^{+}$ & $\lambda$2423             &          & 6.13(-3)\\
Ne$^{4+}$/H$^{+}$ & $\lambda$3426             & 4.28(-4) & 2.37(-4)\\
Ar$^{3+}$/H$^{+}$ & $\lambda\lambda$4711,4740 & 3.37(-6) & 1.84(-6)\\

\hline

\end{tabular}
\end{table}

\begin{table*}
 \begin{center}
 \begin{tabular}{||lll||}
  \hline
Ion & Transition Probabilities & Collision Strengths \\
  \hline
C~{\sc iii} & Keenan et al 1992, Fleming et al 1996 & Keenan et al 1992 \\
C~{\sc iv} & Wiese et al 1966 & Blum \& Gau \& Henry 1977 \\
N~{\sc ii} & Nussbaumer \& Rusca 1979 & Stafford et al 1994 \\
N~{\sc iv} & Nussbaumer \& Storey 1979, Fleming et al 1995 & Mendoza 1983 \\
O~{\sc ii} & Zeippen 1982 & Pradhan 1976 \\
O~{\sc iii} & Nussbaumer \& Storey 1981 & Aggarwal 1983 \\
O~{\sc iv} & Nussbaumer \& Storey 1982 & Zhang et al 1994 \\
Ne~{\sc iii} & Mendoza 1983 & Butler \& Zeippen 1994 \\
Ne~{\sc iv} & Zeippen 1982 & Giles 1981 \\
Ne~{\sc iv} & Froese Fischer \& Saha 1985 & Lennon \& Burke 1994 \\
Ar~{\sc iv} & Mendoza \& Zeippen 1982 & Zeippen et al 1987 \\
  \hline
 \end{tabular}
 \caption{Atomic Data References}
 \label{CELrefs}
 \end{center}
\end{table*}

\subsection{Ionic abundances from ORLs}

Because of the strong power-law dependence of recombination line fluxes ($\propto$ T$_{e}^{-1}$), ORL emission is expected to originate almost entirely from the coolest regions of the knots.  Temperatures measured from O~{\sc ii} recombination lines should be representative of the temperatures of the knot cores, and we therefore derive ORL abundances of heavy elements using temperatures of 500~K for knot J1 and 2500~K for J3, the means of the values derived from I($\lambda$4649)/I($\lambda$4089) and I($\lambda$4075)/I($\lambda$4089).  Helium abundances are determined using temperatures of 4600~K for J1 and 8850~K for J3, from the means of the helium temperature diagnostics described in section~\ref{tempsanddensities}.  We adopt the electron densities derived from [O~{\sc ii}] forbidden lines, 2800 cm$^{-3}$ and 3200 cm$^{-3}$ for J1 and J3, respectively.  The resulting ORL abundances are given in Table~\ref{ORLionicabunds}, while the abundances for individual ions are discussed below.

\subsubsection{He$^{+}$/H$^{+}$ and He$^{2+}$/H$^{+}$}
\label{Heabundssection}

Ionic and total abundances relative to hydrogen derived from helium recombination lines are given in Table~\ref{Heabunds}.  The He$^{+}$/H$^{+}$ abundance was derived from the $\lambda$4471, $\lambda$5876 and $\lambda$6678 lines, averaged with weights of 1:3:1, roughly proportional to their observed intensity ratios.  Effective recombination coefficients were taken from Brocklehurst (1972).  For the three He~{\sc i} lines used, the difference between Brocklehurst's results and the more recent calculations by Smits (1996) is no more than 1.5 per cent.  Case A was assumed for the $\lambda\lambda$4471,5876 triplet lines, and Case B for the $\lambda$6678 singlet line.  Contributions due to collisional excitation from the 2s\,$^3$S metastable level were corrected for using formulae from Kingdon \& Ferland (1995).  At the low temperatures prevailing in J1, collisional excitation effects are negligible, but in J3 they contribute 2.0, 5.2 and 2.5 per cent of the $\lambda$4471, $\lambda$5876 and $\lambda$6678 line fluxes respectively.  The He$^{2+}$/H$^{+}$ abundance was derived from the He~{\sc ii} $\lambda$4686 line, using effective recombination coefficients from Storey \& Hummer (1995).

Several other He~{\sc i} lines were observed in our spectra, and in Table~\ref{Helines} we compare their observed fluxes relative to that of the $\lambda$4471 line to those predicted by Brocklehurst (1972).  The agreement between theory and observations is generally good for the 2p\,$^1$P$^{\rm o}$ - $n$d\,$^1$D and 2p\,$^3$P$^{\rm o}$ - $n$d\,$^3$D series.  However, the intensities of the 2s\,$^1$S - $n$p\,$^1$P$^{\rm o}$ lines relative to $\lambda$4471 are 40-50 per cent lower than predicted.  If this was caused by self-absorption from the metastable 2s\,$^1$S level, then the 2p\,$^1$P$^{\rm o}$ - $n$s\,$^1$S series (lines at 7281, 5047 and 4438 {\AA}) would be enhanced.  Unfortunately this series is not seen in our spectra.  Liu et al (2000) found that in NGC\,6153, the 2s\,$^1$S - $n$p\,$^1$P$^{\rm o}$ series was a factor of 2-3 weaker than predicted, but the expected enhancement of the 2p\,$^1$P$^{\rm o}$ - $n$s\,$^1$S series was not seen.  A similar phenomenon was also found in the galactic bulge PNe M\,1-42 and M\,2-36 (Liu et al 2001).  The predicted intensities assume Case B, where the He~{\sc i} Lyman series is optically thick, and for the 2s\,$^1$S - $n$p\,$^1$P$^{\rm o}$ lines the predicted intensities relative to He~{\sc i} $\lambda$4471 are a factor of 50 lower if Case A is assumed instead, so the observed intensities could be explained by a departure from pure Case B recombination towards case A.  Such a departure could arise through the destruction of He~{\sc i} Lyman series photons by dust.  The core of Abell~30 is known to be very dusty (Cohen \& Barlow 1974, Harrington 1996).  Another possibility is that Case B is inappropriate simply due to the small physical size and hence low optical depth of the knots.

\begin{table}
%\begin{minipage}{120mm}
\centering
\caption{He/H abundance ratios, by number}
\label{Heabunds}
\begin{tabular}{ccccc}

\hline
He$^{i+}$/H$^{+}$ & Line & J1 & J3 \\
\hline

He$^{+}$/H$^{+}$  & He~{\sc i} $\lambda$4471  & 8.560 & 8.766 \\
He$^{+}$/H$^{+}$  & He~{\sc i} $\lambda$5876  & 8.545 & 8.562 \\
He$^{+}$/H$^{+}$  & He~{\sc i} $\lambda$6678  & 8.678 & 8.692 \\
He$^{+}$/H$^{+}$  & Mean                      & 8.575 & 8.629 \\
He$^{2+}$/H$^{+}$ & He~{\sc ii} $\lambda$4686 & 2.203 & 3.015 \\
He/H              &                           & 10.78 & 11.64 \\

\hline

\end{tabular}
\end{table}

\begin{table}
%\begin{minipage}{120mm}
\centering
\caption{Comparison of dereddened He~{\sc i} lines strengths with predictions from Brocklehurst (1972) and Smits (1996) for T$_{e}$=4600 K (J1) and 8850 K (J3)}
\label{Helines}
\begin{tabular}{cccccccc}

\hline
$n$ & $\lambda_{{\rm o}}$ & $I_{{\rm obs}}$ & $I_{\rm pred}$ & $I_{\rm pred}$ & $I_{{\rm obs}}$ & $I_{\rm pred}$ & $I_{\rm pred}$ \\
    &                     & J1              & B72            & S96            & J3              & B72            & S96\\
\hline
 & & & & & & & \\
 \multicolumn{8}{c}{2p\,$^1$P$^{\rm o}$ - $n$d\,$^1$D series} \\
3 & 6678.16 & 0.829 &       & 0.861 & 0.777 & 0.809 & 0.789 \\
4 & 4921.93 & 0.256 &       & 0.269 & 0.270 & 0.273 & 0.267 \\
5 & 4387.93 & 0.126 &       & 0.120 & 0.123 & 0.127 & 0.122 \\
6 & 4143.76 & 0.074 &       &       & 0.135 & 0.071 &       \\
 & & & & & & & \\
 \multicolumn{8}{c}{2p\,$^3$P$^{\rm o}$ - $n$d\,$^3$D series} \\
3 & 5875.66 & 3.060 & 3.051 & 2.997 & 2.760 & 2.820 & 2.734 \\
4 & 4471.50 & 1.000 & 1.000 & 1.000 & 1.000 & 1.000 & 1.000 \\
5 & 4026.21 & 0.571 & 0.456 & 0.448 & 0.523 & 0.470 & 0.462 \\
6 & 3819.62 & *     & 0.249 &       & 0.240 & 0.261 &       \\
7 & 3705.02 & *     & 0.153 &       & 0.208 & 0.161 &       \\
8 & 3634.25 & 0.135 & 0.100 &       & *     & 0.106 &       \\
 & & & & & & & \\
 \multicolumn{8}{c}{2s\,$^1$S - $n$p\,$^1$P$^{\rm o}$ series} \\
3 & 5015.68 & 0.287 &       & 0.488 & 0.267 & 0.555 & 0.551 \\
4 & 3964.73 & 0.142 &       &       & 0.138 & 0.219 &       \\
 & & & & & & & \\
 \multicolumn{8}{c}{2s\,$^3$S - $n$p\,$^3$P series} \\
3 & 3888.65 & 1.469 & 1.854 & 1.809 & 1.587 & 2.178 & 2.162 \\

\hline

\end{tabular}
\end{table}

\subsubsection{{\rm C}$^{2+}$/{\rm H}$^{+}$}

The C$^{2+}$/H$^{+}$ ratio was derived using only the $\lambda$4267 line, taking effective recombination coefficients from Davey et al (1999).  Only one other C~{\sc ii} line ($\lambda$6462) was detected in the spectrum of J3, and no others were seen in J1.  The C$^{2+}$/H$^{+}$ abundances thus derived are colossal (Table~\ref{CIIORLabunds}), nearly half that of hydrogen and several hundred times higher than the value derived from the UV $\lambda\lambda$ 1906,1909 CELs of C~{\sc iii}].  The C$^{2+}$ abundance derived from $\lambda$4267 is not sensitive to case: the recombination coefficient varies by only 1 per cent between Case A and Case B recombination (Davey et al 1999)

To explain the frequently-observed discrepancy between C$^{2+}$ abundances derived from $\lambda$4267 and the $\lambda\lambda$1906,1909 lines, Barker (1982) considered several possible mechanisms for enhancing the $\lambda$4267 line, including charge transfer, dielectronic recombination, blending and resonance fluorescence.  Jacoby \& Ford considered the high carbon abundance they derived for J3 to be questionable in the light of these possibilities, and suggested that it could be too high by a factor of up to 10.

The 3d-4f $\lambda$4267 line is mainly fed by 4f\,$^2$F$^{\rm o}$--ng\,$^2$G transitions.  The other C~{\sc ii} line seen in the spectrum of J3 is the 4f\,$^2$F$^{\rm o}$--6g\,$^2$G $\lambda$6462 line.  The ratio of this line to $\lambda$4267 is predicted by recombination theory (Davey et al 1999) to be 0.104, which is in reasonable agreement with the observed value of 0.089.  It would be useful to observe more high-excitation C~{\sc ii} lines to further confirm the reliability of abundances based on $\lambda$4267, but nonetheless this result indicates that no unknown process in addition to recombination is populating the 4f\,$^2$F$^{\rm o}$ level and leading to an overestimated $\lambda$4267 C$^{2+}$/H$^{+}$ abundance ratio.  4f--ng transitions have also been detected in NGC\,6153, M\,1-42 and M\,2-36 (Liu et al 2000, 2001), and in each case the observed intensities are in very good agreement with the predictions of recombination theory.

\begin{table}
%\begin{minipage}{120mm}
\centering
\caption{C$^{2+}$/H$^{+}$ abundances from ORLs}
\label{CIIORLabunds}
\begin{tabular}{ccccc}

\hline
$\lambda_{\rm 0}$ & \multicolumn{2}{c}{J1} & \multicolumn{2}{c}{J3} \\
 & $I_{obs}$ & C$^{2+}$/H$^{+}$ & $I_{obs}$ & C$^{2+}$/H$^{+}$ \\

\hline

4267.15 & 6.851 & 0.376 & 5.655 & 0.413 \\
6461.95 & *     & *     & 0.502 & 0.361 \\

\hline

\end{tabular}
\end{table}

\subsubsection{{\rm N}$^{2+}$/{\rm H}$^{+}$}

Seven N~{\sc ii} lines were observed from one or both knots, and the N$^{2+}$/H$^{+}$ ionic abundances derived from them are given in Table~\ref{NIIORLabunds}.  Multiplet V3 recombination coefficients are fairly case-insensitive, with the Case B value being about 20 per cent higher.  All the 3d-4f transitions are also case insensitive.  Abundances were derived assuming Case A for singlets and Case B for triplets, using the recombination coefficients of Escalante \& Victor (1990).  The adopted N$^{2+}$/H$^{+}$ abundance ratios are the values found by averaging the results from multiplet V3 with the co-added 3d-4f transition values, and are 0.22 for J1 and 0.20 for J3.

% values derived for Te=500K

\begin{table}
%\begin{minipage}{120mm}
\centering
\caption{N$^{2+}$/H$^{+}$ abundances from ORLs}
\label{NIIORLabunds}
\begin{tabular}{lccccc}

\hline
$\lambda_{\rm 0}$ & Mult & \multicolumn{2}{c}{J1} & \multicolumn{2}{c}{J3} \\
 & & $I_{obs}$ & N$^{2+}$/H$^{+}$ & $I_{obs}$ & N$^{2+}$/H$^{+}$ \\

\hline

5666.63           & V3     &   *       & *       & 0.243   & 0.207 \\
5679.56           & V3     &   0.563   & 0.228   & 0.415   & 0.190 \\
\multicolumn{2}{c}{\bf V\,3 3s\,$^3$P$^{\rm o}$ -- 3p\,$^3$D} &{\bf 1.173}&{\bf 0.23}&{\bf 0.892}&{\bf 0.20}\\
                  &        &           &         &         &       \\
\multicolumn{6}{c}{\bf 3d-4f transitions} \\
4041.31           & V39b   &   0.775   & 0.235   & 0.337   & 0.156 \\
4236.91,4237.05   & V48a   &   0.459   & 0.225   & 0.364   & 0.238 \\
4241.24,.78       & V48b   &   0.525   & 0.210   & *       & *     \\
4442.02           & V55a   &   0.139   & 0.256   & *       & *     \\
4530.41           & V58b   &   0.265   & 0.120   & 0.333   & 0.227 \\
{\bf Sum}         &        &{\bf 2.163}&{\bf 0.20}&{\bf 1.034}&{\bf 0.20}\\

\hline

\end{tabular}
\end{table}

% values derived using Te=4600K

%\begin{table*}
%%\begin{minipage}{120mm}
%\centering
%\caption{N$^{2+}$/H$^{+}$ abundances from ORLs}
%\label{NIIORLabunds}
%\begin{tabular}{cccccc}

%\hline
%$\lambda_{\rm 0}$ & Mult & \multicolumn{2}{c}{J1} & \multicolumn{2}{c}{J3} \\
% & & $I_{obs}$ & N$^{2+}$/H$^{+}$ & $I_{obs}$ & N$^{2+}$/H$^{+}$ \\

%\hline

%5666.63           & V3     &   *       & *       & 0.243   & 0.201 \\
%5679.56           & V3     &   0.563   & 0.251   & 0.415   & 0.185 \\
%\multicolumn{2}{c}{\bf V\,3 3s\,$^3$P$^{\rm o}$ -- 3p\,$^3$D} &{\bf 1.173}&{\bf 0.25}&{\bf 0.892}&{\bf 0.18}\\
%                  &        &           &         &         &       \\
%\multicolumn{6}{c}{\bf 3d-4f transitions} \\
%4041.31           & V39b   &   0.775   & 0.421   & 0.337   & 0.183 \\
%4236.91,4237.05   & V48a   &   0.459   & 0.354   & 0.364   & 0.335 \\
%4241.24,.78       & V48b   &   0.525   & 0.432   & *       & *     \\
%4442.02           & V55a   &   0.139   & 0.460   & *       & *     \\
%4530.41           & V58b   &   0.265   & 0.215   & 0.333   & 0.270 \\
%{\bf Sum}         &        &{\bf 2.163}&{\bf 0.37}&{\bf 1.034}&{\bf 0.24}\\

%\hline

%\end{tabular}
%\end{table*}

\subsubsection{{\rm O}$^{2+}$/{\rm H}$^{+}$}

Many O~{\sc ii} recombination lines are seen from both knots.  The resulting O$^{2+}$/H$^{+}$ ionic abundance ratios are given in Table~\ref{OIIORLabunds}.  The effective recombination coefficients used are from Storey (1994) for 3s-3p transitions, and from Liu et al (1995) for 3p-3d and 3d-4f transitions.  Case A is assumed for doublets and Case B for quartets.  Of the multiplets observed, only V19 and V28 are strongly sensitive to case.  The mean O$^{2+}$/H$^{+}$ abundances derived from these multiplets are (0.8$\pm$0.2) for J1 and (1.0$\pm$0.2) for J3, which are in agreement within the errors with the average values derived from the other 3-3 multiplets, (1.2$\pm$0.2) for J1, and (0.9$\pm$0.2) for J3.

The mean O$^{2+}$/H$^{+}$ abundance ratios derived by averaging with equal weight the values from individual 3-3 multiplets with the co-added 3d-4f values are (1.08$\pm$0.19) for J1 and (0.91$\pm$0.15) for J3.  We adopt these values as the O$^{2+}$/H$^{+}$ recombination line abundance.

% derived using T=500K and 2500K

\begin{table}
%\begin{minipage}{120mm}
\centering
\caption{O$^{2+}$/H$^{+}$ abundances from ORLs}
\label{OIIORLabunds}
\begin{tabular}{lccccc}

\hline
$\lambda_{\rm 0}$ & Mult & \multicolumn{2}{c}{J1} & \multicolumn{2}{c}{J3} \\
 & & $I_{obs}$ & O$^{2+}$/H$^{+}$ & $I_{obs}$ & O$^{2+}$/H$^{+}$ \\

\hline

4638.86 & V1   & 1.153 & 1.234 & 0.689 & 0.696 \\ 
4641.81 & V1   & 2.347 & 0.996 & 1.615 & 0.647 \\ 
4649.13 & V1   & 2.787 & 0.622 & 2.510 & 0.529 \\ 
4650.84 & V1   & 0.172 & 0.184 & 0.737 & 0.745 \\ 
4661.63 & V1   & 1.010 & 0.847 & 0.843 & 0.667 \\ 
4673.73 & V1   & *     & *     & 0.125 & 0.638 \\ 
4676.24 & V1   & 0.713 & 0.712 & 0.641 & 0.604 \\ 
\multicolumn{2}{l}{\bf V\,1 3s\,$^4$P -- 3p\,$^4$D$^{\rm o}$} &{\bf 8.409}&{\bf 0.75}&{\bf 7.232}&{\bf 0.61}\\
4317.14 & V2   & 0.551 & 0.752 & 0.438 & 0.591 \\ 
4319.63 & V2   & 0.412 & 0.520 & 0.213 & 0.266 \\ 
4325.76 & V2   & *     & *     & 0.175 & 1.182 \\ 
4345.56 & V2   & 0.927 & 1.273 & 0.208 & 0.283 \\ 
4349.43 & V2   & 0.815 & 0.444 & 0.740 & 0.399 \\ 
4366.89 & V2   & 1.544 & 1.972 & 0.688 & 0.870 \\ 
\multicolumn{2}{l}{\bf V\,2 3s\,$^4$P -- 3p\,$^4$P$^{\rm o}$} &{\bf 4.578}&{\bf 0.87}&{\bf 2.464}&{\bf 0.49}\\
3749.48 & V3   & 0.807 & 0.926 & 0.609 & 0.651 \\
\multicolumn{2}{l}{\bf V\,3 3p\,$^2$P$^{\rm o}$ -- 3d\,$^2$P} &{\bf 1.620}&{\bf 0.93}&{\bf 1.313}&{\bf 0.70}\\
4414.90 & V5   & 0.662 & 2.381 & *     & *     \\
4416.97 & V5   & 0.464 & 3.013 & 0.426 & 2.009 \\
\multicolumn{2}{l}{\bf V\,5 3s\,$^2$P -- 3p\,$^2$D$^{\rm o}$} &{\bf 1.207}&{\bf 2.66}&{\bf 1.279}&{\bf 2.00}\\
4069.62 & V10  & 0.910 & 0.869 & 0.870 & 0.857 \\ 
4069.89 & V10  & 1.457 & 0.872 & 1.393 & 0.860 \\ 
4072.16 & V10  & 2.448 & 0.967 & 1.871 & 0.763 \\ 
4075.86 & V10  & 2.564 & 0.701 & 1.864 & 0.526 \\ 
4078.84 & V10  & 0.382 & 0.992 & 0.164 & 0.440 \\ 
4085.11 & V10  & 0.378 & 0.799 & 0.243 & 0.531 \\ 
\multicolumn{2}{l}{\bf V\,10 3p\,$^4$D$^{\rm o}$ -- 3d\,$^4$F} &{\bf 8.460}&{\bf 0.83}&{\bf 6.658}&{\bf 0.68}\\
4153.30 & V19  & 0.861 & 0.990 & *     & *     \\ 
4169.22 & V19  & *     & *     & 0.356 & 1.267 \\
\multicolumn{2}{l}{\bf V\,19 3p\,$^4$P$^{\rm o}$ -- 3d\,$^4$P} &{\bf 2.290}&{\bf 0.99}&{\bf 2.781}&{\bf 1.27}\\
4110.78 & V20  & 0.502 & 1.880 & *     & *     \\ 
4119.22 & V20  & 0.787 & 0.798 & 0.475 & 0.507 \\ 
\multicolumn{2}{l}{\bf V\,20 3p\,$^4$P$^{\rm o}$ -- 3d\,$^4$D} &{\bf 2.508}&{\bf 1.03}&{\bf 1.176}&{\bf 0.51}\\
4924.53 & V28  & 0.253 & 0.535 & 0.326 & 0.726 \\ 
\multicolumn{2}{l}{\bf V\,28 3p\,$^4$S$^{\rm o}$ -- 3d\,$^4$P} &{\bf 0.471}&{\bf 0.54}&{\bf 0.607}&{\bf 0.73}\\
        &      &       &       &       &       \\
\multicolumn{6}{c}{{\bf 3d-4f transitions}}\\
4089.29 & V48a & 1.283 & 0.712 & 0.755 & 0.516 \\ 
4083.90 & V48b & 0.418 & 0.817 & 0.240 & 0.573 \\ 
4097.26$^a$ & V48b & 2.893 & 2.431 & 2.183 & 2.298 \\
4087.15 & V48c & 0.404 & 0.826 & 0.185 & 0.465 \\ 
4303.82 & V53a & 1.127 & 1.332 & 1.118 & 1.627 \\ 
4294.78 & V53b & 0.682 & 1.324 & 0.651 & 1.554 \\ 
4281-83 & V53b, V67c & 0.653 & 1.072 & 0.831 & 1.075 \\
4276    & V53c,V67b & 0.772 & 1.365 & 1.270 & 2.764 \\
4291-92 & V55, V78c  & 0.523 & 1.139 & 0.776 & 2.080 \\
4277    & V67c & 1.017 & 2.432 & 0.635 & 1.870 \\
4371.62 & V76b & 0.373 & 2.096 & *     & *     \\ 
4342.00 & V77  & 0.586 & 0.590 & 0.644 & 0.799 \\
4285.69 & V78b & 0.423 & 1.237 & 0.281 & 1.011 \\ 
4491.23 & V86a & *     & *     & 0.274 & 1.363 \\ 
4466.42 & V86b & 0.357 & 1.940 & 0.287 & 1.926 \\ 
4609.44 & V92a & 0.925 & 1.197 & 0.419 & 0.667 \\ 
4602.13 & V92b & *     & *     & 0.690 & 2.760 \\ 
%4060.60 & V97  & 0.744 &       & 0.482 &       \\ No atomic data for this line...?
{\bf sum} &   & {\bf 9.543} & {\bf 1.10}  & {\bf 9.056} & {\bf 1.18 } \\
\hline

\end{tabular}

\begin{description}
\item $^a$ Excluded from co-added 3d--4f intensity: affected by blending with N~{\sc ii} lines excited by secondary Bowen flourescencing.
\end{description}

\end{table}

\subsubsection{{\rm Ne}$^{2+}$/{\rm H}$^{+}$}

The Ne$^{2+}$/H$^{+}$ abundance ratios presented in Table~\ref{NeIIORLabunds} are derived using the line fluxes of the V1 multiplet, using atomic data from Kisielius et al (1998).  The abundances derived are not case-sensitive: the recombination coefficient differs by less than 1 per cent between Case A \& Case B.

\begin{table}
%\begin{minipage}{120mm}
\centering
\caption{Ne$^{2+}$/H$^{+}$ abundances from ORLs}
\label{NeIIORLabunds}
\begin{tabular}{cccccc}

\hline
$\lambda_{\rm 0}$ & \multicolumn{2}{c}{J1} & \multicolumn{2}{c}{J3} \\
 & $I_{obs}$ & Ne$^{2+}$/H$^{+}$ & $I_{obs}$ & Ne$^{2+}$/H$^{+}$ \\

\hline

3694 & 0.665  & 0.174  & *      & *      \\
3664 & 0.718  & 0.439  & 1.390  & 0.954  \\
3777 & 0.332  & 0.219  & *      & *      \\
3710 & *      & *      & 0.526  & 0.390  \\

{\bf V\,1 3s\,$^4$P -- 3p\,$^4$P$^{\rm o}$} & {\bf 2.684} & {\bf 0.25} & {\bf 6.630} & {\bf 0.71} \\
\hline
\end{tabular}
\end{table}

% calculated using T=4600 and 8850K
%
%\begin{table}
%%\begin{minipage}{120mm}
%\centering
%\caption{Ne$^{2+}$/H$^{+}$ abundances from ORLs}
%\label{NeIIORLabunds}
%\begin{tabular}{cccccc}
%
%\hline
%$\lambda_{\rm 0}$ & \multicolumn{2}{c}{J1} & \multicolumn{2}{c}{J3} \\
% & $I_{obs}$ & Ne$^{2+}$/H$^{+}$ & $I_{obs}$ & Ne$^{2+}$/H$^{+}$ \\
%
%\hline
%
%3694 & 0.665  & 0.204  & *      & *      \\
%3664 & 0.718  & 0.51   & 1.39   & 0.971  \\
%3777 & 0.332  & 0.262  & *      & *      \\
%3710 & *      & *      & 0.526  & 0.402  \\
%
%{\bf V\,1 3s\,$^4$P -- 3p\,$^4$P$^{\rm o}$} & {\bf 2.684} & {\bf 0.29} & {\bf 6.630} & {\bf 0.71} \\
%\hline
%\end{tabular}
%\end{table}

\subsubsection{Higher ionisation species}

Several optical recombination lines due to triply ionised species are seen in our spectra, as well as one line ($\lambda$4658) due to quadruply-ionised carbon.  We derive C$^{3+}$/H$^{+}$ ratios from $\lambda$4650 (V1) and $\lambda$4187 (V18) lines strengths, and the N$^{3+}$/H$^{+}$ ratio from $\lambda$4379 (V18).  The results are given in Table~\ref{CNIIIORLabunds}.  Effective recombination coefficients for all three ions were taken from P\'equignot et al (1991).  For both knots, we adopt the mean of the two abundances as the C$^{3+}$/H$^{+}$ recombination line abundance.

\begin{table}
%\begin{minipage}{120mm}
\centering
\caption{C$^{3+}$/H$^{+}$, C$^{4+}$/H$^{+}$ and N$^{3+}$/H$^{+}$ abundances from ORLs}
\label{CNIIIORLabunds}
\begin{tabular}{cccccc}

\hline
& & \multicolumn{2}{c}{J1} & \multicolumn{2}{c}{J3} \\
\hline

$\lambda_{\rm o}$ & Mult & I$_{obs}$ & C$^{3+}$/H$^{+}$ & I$_{obs}$ & C$^{3+}$/H$^{+}$\\
                  &      &           & ($\times 10^{-2}$)&          & ($\times 10^{-2}$)\\
4187 & V18 & 0.323 & 3.033 & 0.291 & 4.706 \\
4649 & V1  & 0.517 & 8.797 & 0.384 & 3.396 \\
     &     &       &       &       &       \\

$\lambda_{\rm o}$ & Mult & I$_{obs}$ & C$^{4+}$/H$^{+}$ & I$_{obs}$ & C$^{4+}$/H$^{+}$\\
                  &      &           & ($\times 10^{-3}$)&          & ($\times 10^{-3}$)\\
4658 & V8  & 0.327 & 4.358 & 0.127 & 2.200 \\
     &     &       &       &       &       \\

$\lambda_{\rm o}$ & Mult & I$_{obs}$ & N$^{3+}$/H$^{+}$ & I$_{obs}$ & N$^{3+}$/H$^{+}$\\
                  &      &           & ($\times 10^{-2}$)&          & ($\times 10^{-2}$)\\
4379 & V18 & 2.615 & 6.411 & 1.958 & 6.237 \\

\hline

\end{tabular}
\end{table}

% Calculations for T=4600 and 8850K

%\begin{table}
%%\begin{minipage}{120mm}
%\centering
%\caption{C$^{3+}$/H$^{+}$, C$^{4+}$/H$^{+}$ and N$^{3+}$/H$^{+}$ abundances from ORLs}
%\label{CNIIIORLabunds}
%\begin{tabular}{cccccc}
%
%\hline
%& & \multicolumn{2}{c}{J1} & \multicolumn{2}{c}{J3} \\
%\hline
%
%$\lambda_{\rm o}$ & Mult & I$_{obs}$ & C$^{3+}$/H$^{+}$ & I$_{obs}$ & C$^{3+}$/H$^{+}$\\
%                  &      &           & ($\times 10^{-2}$)&          & ($\times 10^{-3}$)\\
%4187 & V18 & 0.323 & 4.485 & 0.291 & 4.706 \\
%4649 & V1  & 0.517 & 5.434 & 0.384 & 3.396 \\
%     &     &       &       &       &       \\
%
%$\lambda_{\rm o}$ & Mult & I$_{obs}$ & C$^{4+}$/H$^{+}$ & I$_{obs}$ & C$^{4+}$/H$^{+}$\\
%                  &      &           & ($\times 10^{-3}$)&          & ($\times 10^{-3}$)\\
%4658 & V8  & 0.327 & 6.366 & 0.127 & 2.819 \\
%     &     &       &       &       &       \\
%
%$\lambda_{\rm o}$ & Mult & I$_{obs}$ & N$^{3+}$/H$^{+}$ & I$_{obs}$ & N$^{3+}$/H$^{+}$\\
%                  &      &           & ($\times 10^{-2}$)&          & ($\times 10^{-3}$)\\
%4379 & V18 & 2.615 & 9.481 & 1.958 & 8.268 \\
%
%\hline
%
%\end{tabular}
%\end{table}

Our adopted ORL ionic abundances for C, N, O and Ne ions are summarised in Table~\ref{ORLionicabunds}.

\begin{table}
%\begin{minipage}{120mm}
\centering
\caption{Ionic abundances from ORLs, by number}
\label{ORLionicabunds}
\begin{tabular}{ccc}

\hline
X$^{i+}$/H$^{+}$ & J1 & J3 \\
\hline

He$^{+}$/H$^{+}$  & 8.57     & 8.62    \\
He$^{2+}$/H$^{+}$ & 2.20     & 3.02    \\
C$^{2+}$/H$^{+}$  & 0.38     & 0.41    \\
C$^{3+}$/H$^{+}$  & 5.92(-2) & 4.05(-2)\\
C$^{4+}$/H$^{+}$  & 4.36(-3) & 2.20(-3)\\
N$^{2+}$/H$^{+}$  & 0.22     & 0.20    \\
N$^{3+}$/H$^{+}$  & 6.41(-2) & 6.24(-2)\\
O$^{2+}$/H$^{+}$  & 1.08     & 0.91    \\
Ne$^{2+}$/H$^{+}$ & 0.25     & 0.71    \\

\hline

\end{tabular}
\end{table}

\subsection{Abundances in knot J4}

From CELs observed in the FOS spectra of knot J4, we determined abundances of C$^{2+}$, C$^{3+}$, N$^{+}$, O$^{3+}$ and Ne$^{3+}$ relative to helium.  We assumed that the temperature, density and He$^{2+}$/He$^{+}$ ratio in this knot were the same as those found for knot J3 (sections~\ref{tempsanddensities}, \ref{Heabundssection}).  Guerrero \& Manchado (1996) found that the electron densities in knots J2 \& J4 are an order of magnitude lower than those in J3 and J4, but the abundances derived are fairly insensitive to the adopted density, changing by only 10 per cent at most between the two values.  The ionic abundances found for knot J4 are a factor of 3-7 higher relative to helium than those found for knot J3 (Table~\ref{CELionicabunds}), and are shown in Table~\ref{J4abunds}.  This agrees with previous analyses which have shown a chemical segregation between the polar knots and the equatorial ring (Guerrero \& Manchado 1996, Jacoby \& Ford 1983).

\begin{table}
%\begin{minipage}{120mm}
\centering
\caption{Abundances derived from UV lines in knot J4}
\label{J4abunds}
\begin{tabular}{lr}

\hline
Line & X$^{i+}$/He \\
\hline
C$^{2+}$ 1906/8  & 5.2(-4) \\
C$^{3+}$ 1548/51 & 2.6(-4) \\
O$^{3+}$ 1401/4  & 1.4(-2) \\
Ne$^{3+}$ 2423   & 2.8(-3) \\
\hline
\end{tabular}
\end{table}

\subsection{Total elemental abundances}

Total elemental abundances derived from ORLs and CELs are given in Table~\ref{totalabunds}.  They were calculated from the ionic abundances using the ionisation correction scheme of Kingsburgh \& Barlow (1994) to correct for unseen ionisation stages.  The scale is logarithmic, with log N(H) set to 12.

No O$^{+}$/H$^{+}$ abundance is available from ORLs.  To derive a total O/H abundance, we assume that the O$^{2+}$/O$^{+}$ ratio derived from CELs is applicable to ORLs.

For nitrogen, the CEL abundance in J1 is derived from the N$^{+}$/H$^{+}$ abundance.  The ionisation correction factor is given by O/O$^{+}$, which has a value of 11.63.  For J3, the abundance is derived by assuming that N$^{2+}$/H$^{+}$ is half way between N$^{+}$/H$^{+}$ and N$^{3+}$/H$^{+}$.

Only N$^{2+}$ and N$^{3+}$ ORLs are seen, and to derive a total N/H abundance from ORLs, we assume that N$^{+}$/N = O$^{+}$/O.  This correction amounts to only a few percent.

\begin{table}
\caption{Elemental Abundances in units such that ${\rm log} N({\rm H})=12.0$}
\label{totalabunds}
\centering
\begin{tabular}{lcccc}
& \multicolumn{2}{c}{J1} & \multicolumn{2}{c}{J3} \\
\hline
Ion & ORLs & CELs & ORLs  & CELs \\
\hline
He & 13.03 &      & 13.07 &      \\
C  & 11.65 &      & 11.66 & 9.22 \\
N  & 11.49 & 8.88 & 11.43 & 8.90 \\
O  & 12.15 & 9.26 & 12.10 & 9.32 \\
Ne & 11.51 & 9.70 & 11.99 & 9.78 \\
Ar &       & 7.45 &       & 7.22 \\
\hline
\end{tabular}
\end{table}

\section{Discussion}

\subsection{Knots with cool, CNO-rich cores}

The extremely low temperatures measured from oxygen recombination lines, together with the extremely high heavy element abundances relative to hydrogen, show that the two polar knots J1 and J3 analysed here must contain some very cool but still ionised material.  Virtually all the flux from recombination lines would originate in this cool core, while the forbidden line flux would originate in the outer parts of the knot, where the hydrogen-poor material may be mixing with the hydrogen-rich material of the nebula.  The very low temperatures in the core of the knots are assumed to result from the strong cooling due to the enhanced heavy elements.  Preliminary photoionisation modelling using the current observations data of the knots of Abell~30 has shown that models containing a cool ionised core surrounded by a warmer shell can successfully reproduce most of the observed spectrum, although the abundances required do not correspond closely to those determined empirically (Ercolano et al 2002, in preparation).

\subsection{C/O and N/O ratios}

Previous determinations of PN abundances from both ORLs and CELs have found that the discrepancy factor between ORL and CEL abundances is the same for each element (Liu et al 2000, 2001).  Thus, C/O and N/O ionic ratios, as long as both ionic abundances are derived from one type of line or the other, should be reliable.  For Abell~30's knots the discrepancies are broadly similar for C, N and O, ranging from 2.44 dex to 2.89 dex -- factors of 300-700.  Neon has rather lower discrepancies, the ORL abundance being higher than the CEL value by a factor of 65 for J1 and 160 for J3.  This difference may be caused by the weakness of the lines measured and the small number observed.

An important result is that the C/O ratio in the knots is less than unity, whether derived from ORLs or CELs.  The values derived are 0.29 for J1 from CELs, and 0.36 and 0.79 for J3, measured from CELs and ORLs respectively.  Cool, ionised knots similar to those described here are proposed to exist in the planetary nebulae NGC\,6153, M\,1-42 and M\,2-36 (Liu et al 2000, 2001), and for these nebulae the C/O ratios derived from ORLs are also less than unity, at 0.6, 0.15 and 0.76 respectively, implying that their knots could have a similar origin to those in Abell~30.

\subsection{Origin of the knots}

The common explanation for the knots found in Abell~30 and similar objects is the 'Born Again' scenario, in which the central star of a PN which has reached the white dwarf stage experiences a very late helium flash, causing a further ejection of highly processed material into the old planetary nebula (Iben et al 1983).  Abell~58 is another hydrogen-deficient PN, morphologically quite similar to Abell~30, which contains a hydrogen deficient knot (Guerrero \& Manchado 1996), and in this case the knot was ejected in a nova-like explosion in 1916, presumed to be such a late helium flash.  There are some difficulties with the born-again scenario, however: the theory predicts that the ejected material will be carbon-rich (Iben et al 1983), whereas in the case of Abell~30 the C/O ratio is 0.8 for both knots.

Borkowski et al (1993) found from their Hubble Space Telescope images that the two polar knots, J1 and J3, are collinear with the central star to within 5 arcmin.  Such a degree of collimation is hard to explain in a single-star scenario, and Harrington (1996) suggested that the knots may result from a bipolar jet, suggesting an accretion disk within a binary system.

It may be that some hydrogen-deficient PNe are related to classical novae.  DQ Her is an old nova whose shell has many properties in common with the knots of Abell~30: it shows strong recombination line emission, and the strength of the Balmer jump implies a temperature of only 500~K (Williams et al 1978).  Abundance determinations from recombination lines by Petitjean et al (1990) show that heavy element abundances in DQ Her's ejecta are very high relative to helium, and its C/O ratio is 0.36.

\subsection{Implications for other nebulae}

Our confirmation that very cold ionised regions can exist within planetary nebulae is very important.  For some nebulae showing extreme discrepancies between ORL and CEL abundances, temperatures measured from helium and oxygen recombination lines follow the same patterns seen here (Liu 2002), and this strongly suggests that cold ionised metal-rich regions also exist in these nebulae, explaining the differences between observed ORL and CEL abundances as well as the different temperatures from different diagnostics.  These discrepancies, rather than implying large uncertainties in abundance determinations, may instead reflect the fact that CELs and ORLs sample different parts of a nebula, with very different physical conditions.

\end{document}